\documentclass[aps,prb,superscriptaddress,nobibnotes,nofootinbib]{revtex4-2}
\usepackage{graphicx,amsmath,amsfonts,bbold,braket} 
\usepackage{color,bm}
\usepackage{hyperref}
\usepackage{natbib}

\begin{document}
\title{Surface band characters of Weyl semimetal candidate material MoTe$_2$ revealed by one-step ARPES theory}

\author{Ryota Ono}
\affiliation{Graduate School of Science and Engineering, Chiba University, Inage-ku, Chiba-shi 265-8522, Japan}
\author{Alberto Marmodoro}
\affiliation{FZU - Institute of Physics of the Czech Academy of Sciences, Cukrovarnicka 10, CZ-162 53 Prague, Czech Republic}
\author{Jakub Schusser}
\affiliation{New Technologies - Research Center, University of West Bohemia, Univerzitni 8, 306 14 Plze\v{n}, Czech Republic}
\affiliation{Experimentelle Physik VII, Universit\"at W\"urzburg, Am Hubland, D-97074 W\"urzburg, Germany}
\author{Yositaka Nakata}
\affiliation{Graduate School of Science and Engineering, Chiba University, Inage-ku, Chiba-shi 265-8522, Japan}
\author{Eike F. Schwier}
\affiliation{Hiroshima Synchrotron Radiation Center, Hiroshima University, Kagamiyama 2-313, Higashi-Hiroshima 739-0046, Japan}
\affiliation{Experimentelle Physik VII, Universit\"at W\"urzburg, Am Hubland, D-97074 W\"urzburg, Germany}
\author{J\"urgen Braun}
\affiliation{Department of Chemistry, Ludwig Maximililans University M\"unchen, Butenandtstra\ss e 11, 81377 M\"unchen, Germany}
\author{Hubert Ebert}
\affiliation{Department of Chemistry, Ludwig Maximililans University M\"unchen, Butenandtstra\ss e 11, 81377 M\"unchen, Germany}
\author{J\'{a}n Min\'{a}r}
\affiliation{New Technologies - Research Center, University of West Bohemia, Univerzitni 8, 306 14 Plze\v{n}, Czech Republic}
\author{Kazuyuki Sakamoto}
\affiliation{Graduate School of Engineering and Molecular Chirality Research Center, Chiba University, Chiba 263-8522, Japan}
\affiliation{Department of Applied Physics, Osaka University, Osaka 565-0871, Japan}
\author{Peter Kr\"uger}
\email{pkruger@chiba-u.jp}
\affiliation{Graduate School of Engineering and Molecular Chirality Research Center, Chiba University, Chiba 263-8522 Japan}

\date{\today}

\date{\today}
\begin{abstract}
The layered 2D-material MoTe$_2$ in the T$_d$ crystal phase is a
semimetal which has theoretically been predicted to possess
topologically non-trivial bands corresponding to Weyl fermions.  Clear
experimental evidence by angle-resolved photoemission spectroscopy
(ARPES) is, however, lacking, which calls for a careful examination of
the relation between ground state band structure calculations and
ARPES intensity plots.  Here we report a study of the near
Fermi-energy band structure of MoTe$_2$(T$_d$) by means of ARPES
measurements, density functional theory, and one-step-model ARPES
calculations. Good agreement between theory and experiment is
obtained. We analyze the orbital character of the surface bands and
its relation to the ARPES polarization dependence.
We find that light polarization has a major effect on which
bands can be observed by ARPES. For $s$-polarized
light, the ARPES intensity is dominated by subsurface Mo~$d$
orbitals, while $p$-polarized light reveals the bands composed mainly derived from
Te~$p$ orbitals. Suitable light polarization for observing either
electron or hole pocket are determined.
\end{abstract}

\maketitle
\newpage
\section{\label{sec:Intro} Introduction}
Topological Weyl semimetals (WSMs) are receiving much attention because
the quasiparticles at particular points in the band structure are a realization of mass-less Dirac fermions, 
so-called Weyl fermions~\cite{Xu2015}. WSMs have peculiar transport properties \cite{Aji2012,Zyuzin2012,Son2013,Burkov2018},
including the quantum anomalous Hall effect, and the violation of separate number conservation laws for the left
handed and right handed Weyl fermion in the presence of parallel electric and magnetic fields,
known as chiral anomaly~\cite{Nielsen1983}. 
In a semi-metal, both the conduction and the valence bands cross the
Fermi energy. In a WSM, there are special crossing points, called Weyl points, which
correspond to topologically protected states. Weyl points come in pairs with opposite chirality.
They are the end-points of the Fermi arc, i.e. the open Fermi line of the surface band structure.
TaAs and related compounds were the first materials in which WSM property was experimentally
observed~\cite{Weng2015,Xu2015, Lv2015a, Huang2015}. 
More recently, another possible realization of WSMs was found in transition metal dichalcogenides (TMDC), e.g. WTe$_2$~\cite{Soluyanov2015a,Wang2016,Chang2016,Fanciulli2020}.
TMDC are 2D-layered materials and particularly promising for electronic applications.
TaAs and WTe$_2$ are classified as type-I and type-I\hspace{-.1em}I WSMs, respectively~\cite{Soluyanov2015a}.
Both types have point-like crossings at the Fermi energy, but in type-I\hspace{-.1em}I WSM the cone-shaped bands are tilted in k-space.
Also, type-I WSM respects Lorentz invariance, whereas type-I\hspace{-.1em}I breaks it \cite{Soluyanov2015a}.

The TDMC molybdenum tellurite, in the low temperature MoTe$_2$(T$_d$) phase,
has been proposed as a possible WSM material \cite{Sun2015,Crepaldi2017,Jiang2016,Tamai2016}.
At room temperature, MoTe$_2$ crystallizes in the monoclinic, centrosymmetric 1T'
phase. Upon cooling below 240~K MoTe$_2$ changes to the orthorhombic,
non-centrosymmetric T$_d$ phase (space group $Pmn2_1$, No. 31) \cite{Brown1966,Hughes1978,Zandt2007,Qi2016a}.
While the atomic structures of the 1T' and T$_d$ phases are similar,
the fact that the T$_d$-phase lacks inversion symmetry makes it a
possible realization of type-I\hspace{-.1em}I WSM, as proposed both on
theoretical and experimental grounds
\cite{Soluyanov2015a,Sun2015,Crepaldi2017,Jiang2016,Tamai2016}.
The surface band structure of MoTe$_2$(T$_d$) has been studied
using ARPES by several authors~\cite{Tamai2016,Jiang2016,Crepaldi2017,Weber2018}
but the interpretation of the data is difficult without dedicated
ARPES simulations including final state and matrix element effects.
ARPES peak positions are routinely used for band mapping, but the peak intensities,
which contain useful information about the electronic wave functions \cite{Puschnig2009}, is often left unexploited.
The ARPES intensity and its light polarization dependence is determined not only by the initial
state band character, but also by final state effects, in particular in spin-orbit coupled systems
\cite{Datzer2017} such as WSMs. As a consequence, reliable ARPES calculations within the one-step model
of photoemission are necessary for a correct interpretation of the experimental data.
To the best of our knowledge, such calculations have not been reported yet.
Aryal et al. investigated the ARPES intensity using the plane-wave
approximation~\cite{Aryal2019} which has many known shortcomings~\cite{Bradshaw_2015}, especially for
heavy elements like Mo and Te, where distorted wave effects are large.

Most ARPES studies of MoTe$_2$(T$_d$) have focused on the search for
topologically non-trivial Weyl points and Fermi arcs~\cite{Sun2015,Crepaldi2017,Jiang2016,Tamai2016,Weber2018}.
While density functional theory (DFT) calculations have consistently found these
features in the MoTe$_2$ surface band structure, the experimental evidence
remains elusive and controversial. This calls for
a careful examination of the relation between the band dispersion
predicted by ground state DFT and the ARPES
intensity maps that are recorded in experiment. Moreover, the
polarization dependence of ARPES can be utilized for highlighting
different bands, and for revealing their elemental and orbital
character. The orbital character of the surface bands is important
for technological applications, since it determines the
sensitivity of the electronic structure of the system against
controlled and uncontrolled chemical reactions at the surface.

Here we report a detailed study of the surface band structure of
MoTe$_2$(T$_d$) by means of DFT calculations.
To the best of our knowledge, we present the first ARPES calculations
for this system using a one-step model description. We analyze the orbital
character of the near-Fermi level bands and predict a strong
polarization dependence of the ARPES spectra.
The calculated ARPES intensity maps are in good agreement
with new, high-resolution experimental data obtained with a LASER source.
The results show how specific
bands and atomic species can be highlighted in ARPES with appropriate
polarization, providing valuable guidelines for future ARPES measurements.

The rest of the paper is organized as follows.
In Section~\ref{sec: Theoretical}
the computational and experimental methods are outlined.
In Sec.~\ref{sec: results and discussions},
the results of the ground state band structure and of the ARPES calculations
are presented. The polarization dependence of the ARPES spectra is studied
in detail and analyzed in terms of the orbital character of the bands.
Then, the theoretical ARPES maps are compared with experiment.
Finally, in Sec.~\ref{sec: conclusions} we summarize and conclude our work.

\section{\label{sec: Theoretical} Theoretical and experimental methods}
Two types of calculations are performed.
First, the ground state band structure at the MoTe$_2$(T$_d$) surface is computed with DFT in a repeated slab geometry.
Second, ARPES simulations are performed using the one-step model of photoemission.
Throughout this paper, the experimental structure with lattice constants
$a=3.477$~\AA, $b=6.335$~\AA, $c=13.883$~\AA~\cite{Qi2016a} is used.
The crystal structure is shown in Fig.~\ref{fig: struct}.
Surface relaxation has been checked with DFT and found negligible,
as expected for layered materials with weak van-der-Waals forces between
layers such as MoTe$_2$.
\subsection{Ground state band structure calculations}
The surface is modeled using a slab of four MoTe$_2$ layers and over 10~\AA\ vacuum spacing
between slabs, resulting in a supercell lattice parameter $c=37.8$~\AA.
The projector-augmented wave method as implemented in the Vienna
\textit{ab initio} simulation package (VASP) \cite{Kresse1996} is used
with the commonly adopted~\cite{Sun2015} Perdew-Burke-Ernzerhof (PBE)
exchange-correlation functional \cite{Perdew1996}.
The calculations are performed on a $16\times 10\times 1$
\textbf{k}-points mesh in the Brillouin zone and the plane-wave basis
energy cut-off is set to 400 eV. The
spin-orbit interaction (SOI) is taken into account in all calculations.

\subsection{One-step ARPES calculations}
The ARPES calculations are done with the spin-polarized relativistic
Korringa-Kohn-Rostoker (SPR-KKR) package \cite{Ebert2011}.
As in the slab calculations, the experimental crystal structure
and the PBE exchange-correlation potential are used. The
SOI is treated exactly through the Dirac equation.
The Atomic Sphere Approximation (ASA) is used and the KKR equations
are solved with an angular momentum cut-off of $l_{max}=3$.
After computing the bulk Green's function,
a surface model is constructed and the Green's function of the semi-infinite
surface is found by solving a Dyson equation.
We perform one-step model ARPES calculations
\cite{Ebert2011, Braun2013, Minar2011, Braun2018} for the semi-infinite surface model
using the Layered Korringa-Kohn-Rostoker (LKKR) multiple
scattering theory~\cite{Braun_1996, MACA1988381, Kambe1971, Braun2018}.

\begin{figure}[htb]
 \centering
 \includegraphics[width=12cm]{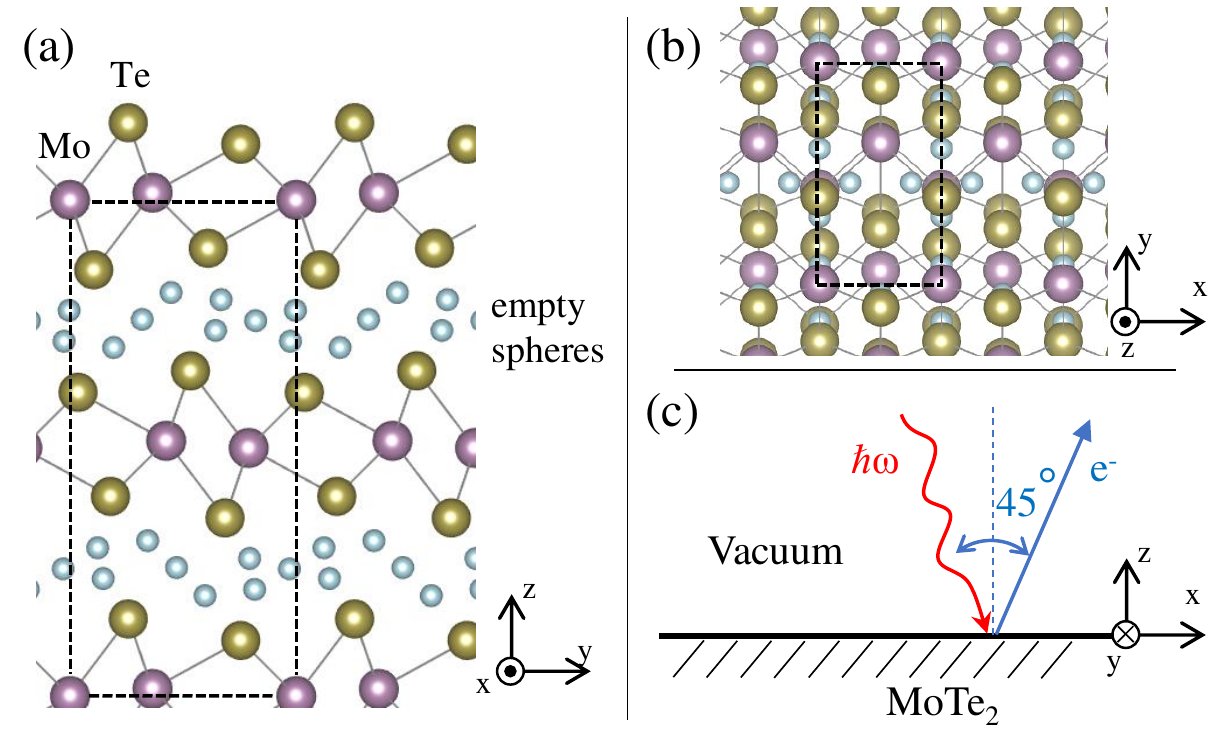} 
\caption{\label{fig: struct}
(a,b) Ball and stick model of MoTe$_2$ surface. Dotted lines indicate unit cell. Mo in purple, Te in yellow, empty spheres in light blue.
(a) Side view. (b) Top view. (c) Experimental geometry.}
\end{figure}

As the crystal structure of MoTe$_2$ is not densely packed but
contains a large interstitial volume, empty spheres must be added when using the ASA
to obtain converged, self-consistent potentials.
As a result, SPR-KKR yields a ground state band structure in excellent agreement 
with the VASP calculations.
However, we found that the empty spheres inside
a MoTe$_2$ layers lead to numerical difficulties in the LKKR calculations, needed for ARPES.
This is related to the fact that in LKKR, compact 2D scattering layers
must be defined. In MoTe$_2$ the crystal planes are rumpled which may lead
to complications when solving the multiple scattering equations between
layers~\cite{Kambe1971, MACA1988381}. Therefore,
in the ARPES calculations, we have removed the empty spheres inside the MoTe$_2$
layers and kept only those between the layers, see Fig.~\ref{fig: struct}~a.
The same strategy has already been applied successfully to
another TMDC material (WTe$_{2}$, Ref.~\cite{Schusser2019, Fanciulli2020}).
As a result, the ARPES band dispersions of MoTe$_2$ agree well with the
DFT-VASP bands as seen in Fig.~\ref{fig: comp}.

\subsection{Experimental details}
The MoTe$_2$ sample was grown by chemical vapor transport method. After preparation
at room temperature, the crystal was in the 1T' phase. During the ARPES measurements,
however, the sample was in the T$_d$-phase, since it was maintained at a temperature of 67 K,
which is well below the Td-transition temperature (240K).
High-resolution ARPES measurements were performed at the $\mu$-ARPES system of Hiroshima Synchrotron Radiation Center (HiSOR), Japan  \cite{Iwasawa2017}.
Photoelectrons, which were excited by a vacuum-ultraviolet laser ($\hbar \omega = 6.27$~eV), were collected by a hemispherical photoelectron analyzer (VG Scienta R4000).
The light was incident in the xz-plane, making an angle of 45$^\circ$ with the electron emission direction (Fig.~\ref{fig: struct}~c).
The energy and spatial resolutions were better than 3 meV and 5$\mu$m, and the angular resolution was better than 0.05$^{\circ}$.

\section{\label{sec: results and discussions} Results and discussions}
\subsection{Band structure of MoTe$_2$(T$_d$)}
\begin{figure}[htb]
\centering
\includegraphics[width=10cm]{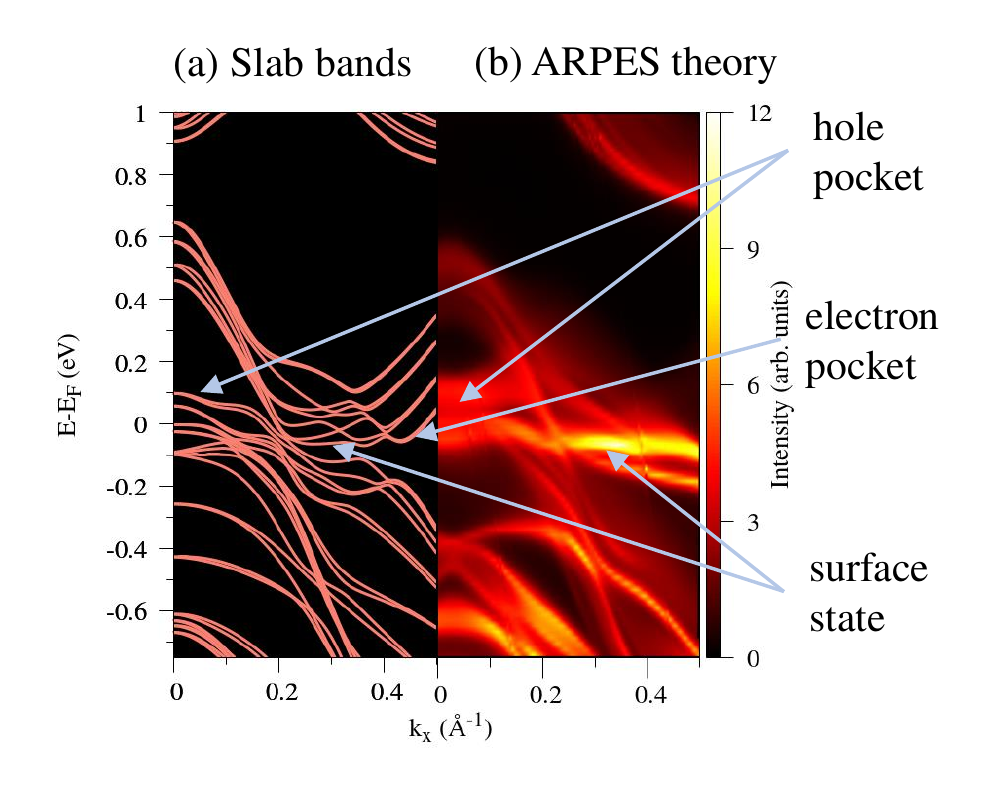}
\caption{\label{fig: comp} Theoretical MoTe$_2$ band structure
  along $\overline{\Gamma}-\overline{X}$ of the 2D Brillouin zone.
  (a) Ground state slab model calculations.
  (b) ARPES calculations for unpolarized light and a photon energy of
  60~eV. 
  }
\end{figure}

In MoTe$_2$(T$_d$), topologically non-trivial bands have been predicted
near the $\overline{\Gamma}-\overline{X}$ line ($k_y=0$) 
of the Brillouin zone between the hole pocket centered
at $k_x$=0 and the electron pocket around $k_x$=0.3~\AA$^{-1}$~\cite{Sun2015}.
Here we focus on this region in (${\bf k}$,$E$) space.
The DFT band structure obtained with the slab model is shown 
in Fig.~\ref{fig: comp} along with ARPES simulations for 
unpolarized light, obtained by summing over x-, y- and z-linear polarization.
Here, we have chosen a photon energy of 60~eV as used in the experiments of Ref.~\cite{Tamai2016}.
The band dispersion in the ARPES map agrees well with the VASP-DFT band
structure, which shows that the ASA used in the KKR calculations
provides an accurate representation of the crystal potential of MoTe$_2$.
The band structure in Fig.~\ref{fig: comp} is
in good agreement with other calculations~\cite{Sun2015} and features
an electron and a hole pocket along~$k_x$, as well as a flat surface state.

\subsection{Orbital projected bands and ARPES polarization dependence}
Here we analyze the orbital character of the bands near the Fermi level and
the relation to the ARPES polarization dependence.
In the one-step ARPES calculations, final state and matrix element effects are fully
taken into account~\cite{Minar2011} and the predicted polarization dependence should be a reliable guide
for experiments. In addition, a semi-quantitative model of the polarization dependence
based on the valence band structure alone is very useful and can
be gained by considering optical selection rules.
The MoTe$_2$ valence band near $E_F$ is dominated by Mo~4$d$ and Te~5$p$ orbitals as seen in
Fig.~S1 of the Supplemental Material~\cite{SI}. We limit our analysis to these two orbitals.
The photoemission intensity
is proportional to the square of the transition matrix element
\begin{equation}
M_{if} = \braket{\phi_f|\hat{\epsilon} \cdot \bm{r}|\phi_i}, 
\end{equation}
where $\ket{\phi_i}$, $\ket{\phi_f}$, $\hat{\epsilon}$ and $\bm{r}$ are initial and final state
wave function, light polarization vector and electron position operator,
respectively.
According to the dipole selection rules, an electron can transit
from an initial state with angular momentum~$l$ to a
final state of angular momentum~$l\pm1$. In the following,
we discuss only $l\rightarrow l-1$ transitions for simplicity.
The $l\rightarrow l+1$ transitions, which often dominate at high energy,
give rise to a more complex, but generally less
pronounced polarization dependence.
To see this, consider e.g. a $p_z$ initial state.
The $l\rightarrow l-1$ transition
leads to an $s$-wave final state and a strong $\cos^2\theta$ polarization
dependence, where $\theta$ is the angle between the electric field vector and the $z$-axis.
In particular, the intensity is zero for $x$- and $y$-linearly polarized light.
In contrast, $l\rightarrow l+1$ transitions from $p_z$ to one of the five $d$-orbitals
are possible for any light polarization such that the intensity never vanishes.

\begin{figure}[htb]
\centering
\includegraphics[width=14cm]{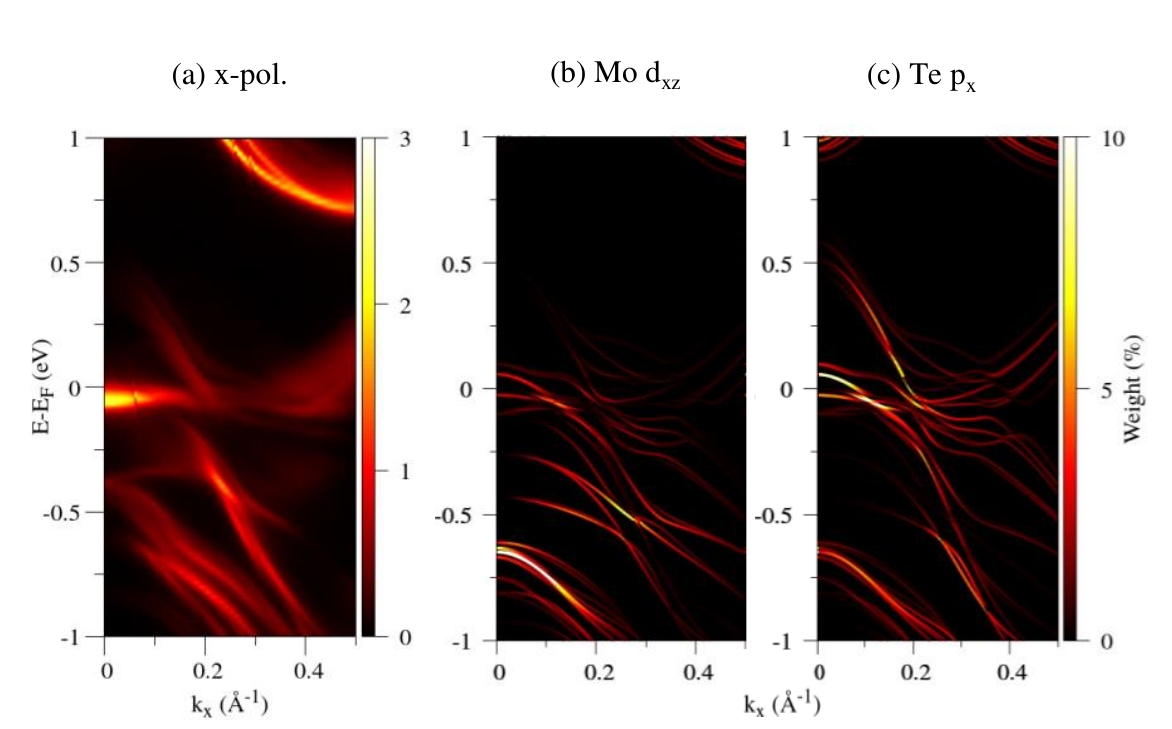}
        \caption{\label{fig:xpol} (a) Calculated ARPES map for x-polarized light ($\hbar \omega = 60$~eV). Note the strong intensity
        at the hole pocket. (b,c) DFT bands projected on the Mo $d_{xz}$ (b) and Te $p_{x}$ orbital (c).}
\end{figure}

In the $l\rightarrow l-1$ channel considered here,
only the following transitions are possible from Mo~$d$ and Te~$p$
initial states for linear polarized light.
X-polarization: Te~$p_x\rightarrow s$, Mo~$d_{xz}\rightarrow p_z$.
Y-polarization: Te~$p_y\rightarrow s$, Mo~$d_{yz}\rightarrow p_z$ and
Mo~$d_{xy}\rightarrow p_x$.
Z-polarization: Te~$p_z\rightarrow s$ and Mo~$d_{z^2}\rightarrow p_z$.
Other transitions are forbidden by the dipole selection rules.
From these considerations, we expect that ARPES with x-polarized light
will reveal the bands with a large Te~$p_x$ and Mo~$d_{xy}$ orbital
character.
The calculated ARPES spectra for 60~eV light and x-polarization
are compared with the DFT bands projected onto Te $p_x$ and Mo $d_{xz}$
orbitals in Fig.~\ref{fig:xpol}.
Here and in all following DFT band plots,
projection is done on the surface Te atoms and the subsurface Mo atoms.
The Mo intensity is divided by a factor of two in order to roughly
account for the ARPES surface sensitivity.
Note that this Mo:Te weight ratio also corresponds to bulk MoTe$_2$.
Most ARPES features can be well identified with either of the two initial
states. DFT bands projected on other Te or Mo orbitals
resemble much less the ARPES map, as can be
seen in Figs.~S2 and S3 of the Supplemental Material~\cite{SI}.
This indicates that the orbital character of the bands together with the
$l\rightarrow l-1$ dipole selection rules provides a qualitative
understanding of the polarization dependence of the ARPES spectra.
In Fig.~\ref{fig:xpol}~b,c,
the strongest intensity appears around $k_x =0$~\AA$^{-1}$
($= \bar{\Gamma}$ in 2D Brillouin Zone) for
both Te~$p_x$ and Mo~$d_{xz}$ initial states.
The ARPES for x-polarization is most intense
around $k_x =0$~\AA$^{-1}$ just below the Fermi level,
which can be attributed to the Te $p_x$ orbital.
In addition, other high intensity ARPES bands correspond to the
hole pocket. As seen from Fig.~\ref{fig:xpol} they are mainly due to
Mo $d_{xz}$ character.
Small energy shifts exist between the KKR-computed ARPES bands
and the VASP-DFT bands. This is expected because of the different
approximations used in the two approaches,
most importantly the atomic sphere approximation
in KKR. With this in mind, the most intense Mo~$d_{xz}$ bands
(at $E\sim -0.7$~eV, $k_x\sim 0$) also agree well with the ARPES bands.
From the foregoing analysis, we conclude that linear polarization along
the x-axis is a good choice for observing
the hole pocket in MoTe$_2$. In experiment, this corresponds to
$s$-polarization with light incidence in the yz plane.

\begin{figure}[htb]
\centering
 \includegraphics[width=18cm]{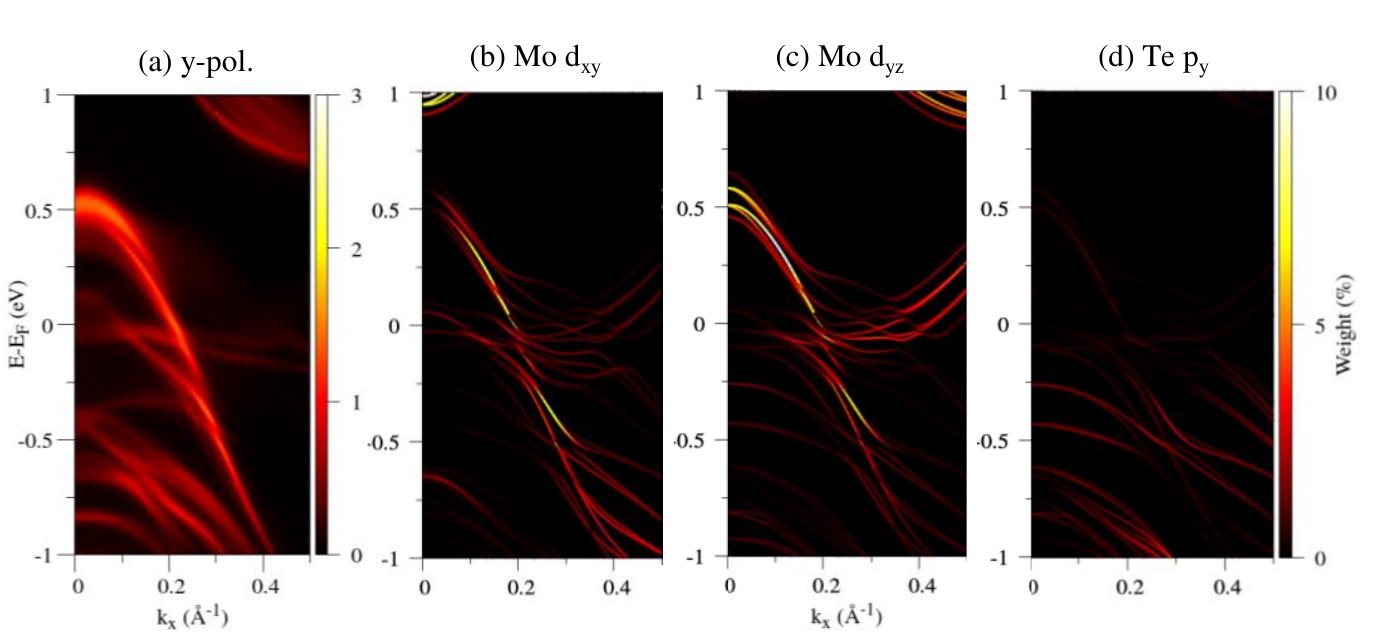}
        \caption{\label{fig:ypol} (a) Calculated ARPES intensity for y-polarized light ($\hbar \omega = 60$~eV). A group of strong ARPES
        bands form a linearly dispersing line. (b-d) DFT bands projected on the Mo $d_{xy}$ (b), Mo $d_{yz}$ (c) and Te $p_{y}$ orbital (c).}
\end{figure}

The calculated ARPES spectra for y-polarization and the
corresponding projected bands are shown in
Fig.~\ref{fig:ypol}. In the case of y-polarization, $l\rightarrow l-1$
transitions are possible only from Mo $d_{xy}$, Mo~$d_{yz}$ and Te~$p_y$ initial states.
In the Mo $d_{xy}$ and Mo $d_{yz}$ projection, the dominant feature is
a group of bands which disperses linearly from
$k_x = 0$~\AA$^{-1}$ $E=0.5$~eV to $k_x=0.4$~\AA$^{-1}$, $E=-1$~eV.
Closer inspection shows that they are made of two groups of bands,
where the upper part joins the electron pocket at $k_x\approx 0.3$~\AA$^{-1}$.
The lower part evolves into the hole pocket around the $\bar{\Gamma}$ point.
These linearly dispersing bands are clearly seen
as a bright feature in the calculated ARPES spectra.
The projection on the Te~$p_z$ bands shows moderately intense bands
in the region $k_x<0.3$~\AA$^{-1}$, $E<-0.7$~eV which can explain the
corresponding bands seen in the ARPES plot.
However, the Te $p_y$ contribution is very weak for all bands
above $-0.4$~eV, and so the Te $p_y$ orbital plays no role for the
electron and hole pockets.

For both $x$- and $y$-polarized light, the calculated ARPES intensity is
very weak compared to $z$-polarization (see Fig.~\ref{fig:zpol} below).
This is because
the considered $k$-range around the $\bar{\Gamma}$ point of the Brillouin zone
corresponds to near normal emission, i.e. an emission direction
perpendicular to the photon electric field vector in the $xy$-plane.
It is well known that
perpendicular emission is generally much weaker than parallel
emission~\cite{Goldberg1981}.
In the popular plane wave
approximation, perpendicular emission is even
impossible~\cite{Puschnig2009}.
Yet, the existence of some pronounced
ARPES features at $\bm{k}=0$ for both $x$- and $y$-polarized light shows
that the plane wave approximation can be
misleading~\cite{Bradshaw_2015,Kruger2018} and that one-step model calculations
are needed for a correct interpretation of ARPES spectra.

In Fig.~\ref{fig:zpol} the calculated ARPES spectra for z-polarization
are shown along with possible orbital projected DFT bands,
namely Mo~$d_{z^2}$ and Te~$p_z$.
The Mo $d_{z^2}$-projection shows strong intensity for a group of
bands which form an arc-like structure around $E=-0.8$~eV
which agrees well with the most intense features of the ARPES map.
Concerning the Te~$p_z$ projected bands, large intensity is seen
for a flat band around the electron pocket,
which has been identified as a surface state~\cite{Sun2015,Crepaldi2017}.
This corresponds to a bright flat line in the ARPES map at $k_x>0.25$~\AA$^{-1}$, $E\approx -0.1$~eV.
By comparison with Figs.~\ref{fig:xpol},\ref{fig:ypol} it is clear that
electron pocket and surface state can be best visualized with $z$-polarized light.
This is in agreement with the experiments by Crepaldi et al.~\cite{Crepaldi2017}
who observed the surface state with strong intensity with mixed $s$- and $p$-polarized 
light (containing some $z$-polarization)
while the surface state intensity almost vanished for pure $s$-polarized light
(which contains zero $z$-polarization).
In $z$-polarization,
the electric field vector of the photon field is nearly parallel to
the emission direction.
This explains why the overall ARPES intensity is much larger
than for in-plane ($x$ or $y$) polarization.
Experimentally, exact $z$-polarization is impossible
but can be approached by using $p$-polarization and grazing incidence.

\begin{figure}[htb]
\centering
 \includegraphics[width=14cm]{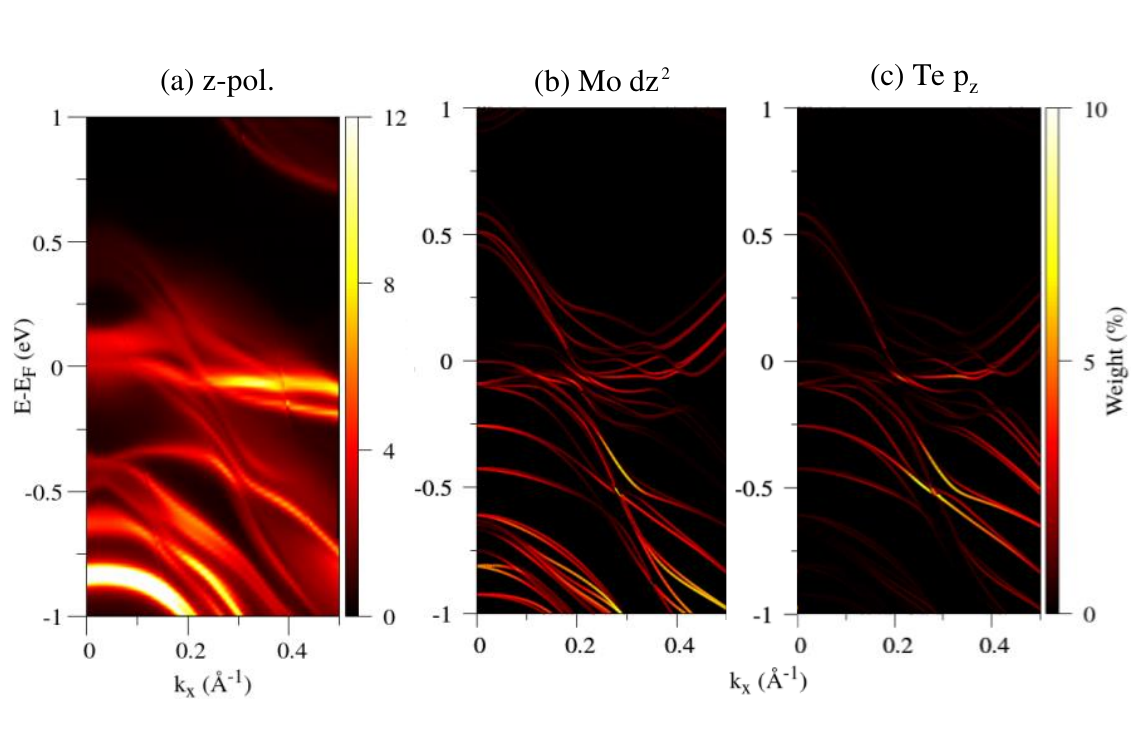}
\caption{\label{fig:zpol}
  (a) Calculated ARPES intensity for z-polarized light ($\hbar \omega = 60$~eV).
  Note that the color scale is enhanced by a factor of~4 as compared
  to Figs.~\ref{fig:xpol} and \ref{fig:ypol}.
  (b,c) DFT bands projected on the Mo $d_{z^2}$ (b) and
  Te $p_{z}$ orbital (c). }
\end{figure}

We note that
Aryal et al.~\cite{Aryal2019} have analyzed the ARPES polarization dependence of MoTe$_2$
by using the plane-wave approximation
and assuming that the final state wave function does not depend on $z$. Thereby
they obtained the same selection rules as we did
for Te~$p$ initial states, but somewhat different ones for Mo~$d$ initial
states.
For example they predict that with $y$-polarized light there are no
transitions from Mo~$d_{yz}$ initial states.
However, the comparison between Fig. 4(a) and (c) strongly indicates that
such transitions have a large
oscillator strength along the $\bar{\Gamma}$-$\bar{X}$ line.

The foregoing analysis shows that the ARPES polarization dependence
can qualitatively be understood from the orbital-projected DFT bands
and the dipole selection rules.
It is clear however, that such an initial state theory cannot give a
quantitative description of ARPES intensity maps. Indeed, final state
effects play an important role too, and are responsible for the
photon energy dependence. Note that the present one-step-model calculations
include all matrix and final state effects.

Despite its qualitative nature, the analysis based on orbital-projected bands and selection
rules, is very useful for determining what polarization
is best suited for probing particular parts of the band structure
and for understanding the nature of the bonding.
In the case of MoTe$_2$, we have seen that
the electron pocket and the surface state can be observed with z-polarization, whereby
the Te~$p_z$ contribution is highlighted.
X-polarization reveals the Te~$p_x$ bands and is a good choice for observing
the hole pocket. With $y$-polarization,
mainly Mo~$d_{xy}$ and $d_{yz}$ orbitals are probed
which form a group of linear dispersing bands which
join either the electron or the hole pocket.
%

We also computed the ARPES intensity maps in the whole 2D Brillouin zone at the Fermi energy (see Fig.~S4 of the Supplemental 
Material~\cite{SI}).
Concerning the polarization dependence, the same conclusion is reached as in the band structure analysis above,
namely that the electron and hole pockets are best observed with $z$ and $x$ polarized light, respectively.
The spin-orbit coupling in MoTe$_2$ gives rise to a spin splitting of bands of the order of 0.5~eV~\cite{Crepaldi2017}. 
Our one-step model calculations (Fig.~S5 of the Supplemental Material~\cite{SI}) 
show that this brings about a substantial spin-polarization of the photoelectrons,
which is especially pronounced at the surface state (see Fig.~2) when $z$-polarized light is used.

\subsection{Comparison with experiment}
Tamai et al.~\cite{Tamai2016} measured the MoTe$_2$ ARPES along
$k_x$ with 60~eV light and $p$-polarization, corresponding to a combination of 
$x$- and $z$-polarized light.
Both electron and hole pocket were observed with high intensity.
In contrast, $s$-polarization suppresses the electron pocket and the surface state
intensity as observed by Crepaldi et al.~\cite{Crepaldi2017}.
Both experimental results can be explained by our polarization analysis.

\begin{figure}[htb]
\centering
\begin{tabular}{cc}
 \includegraphics[width=13cm]{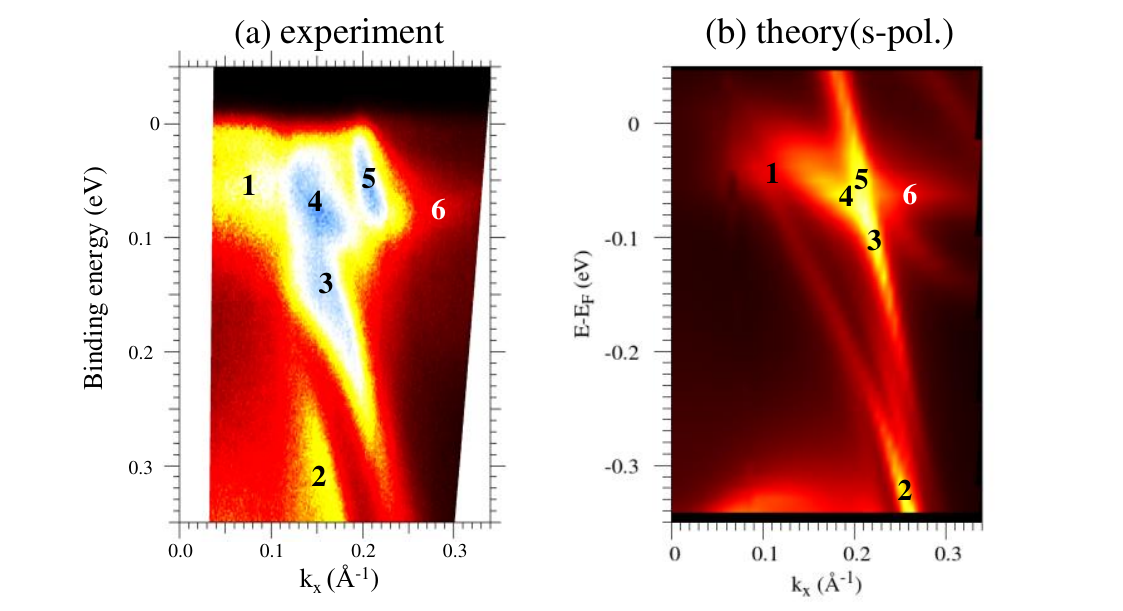}
\end{tabular}

 \includegraphics[width=13cm]{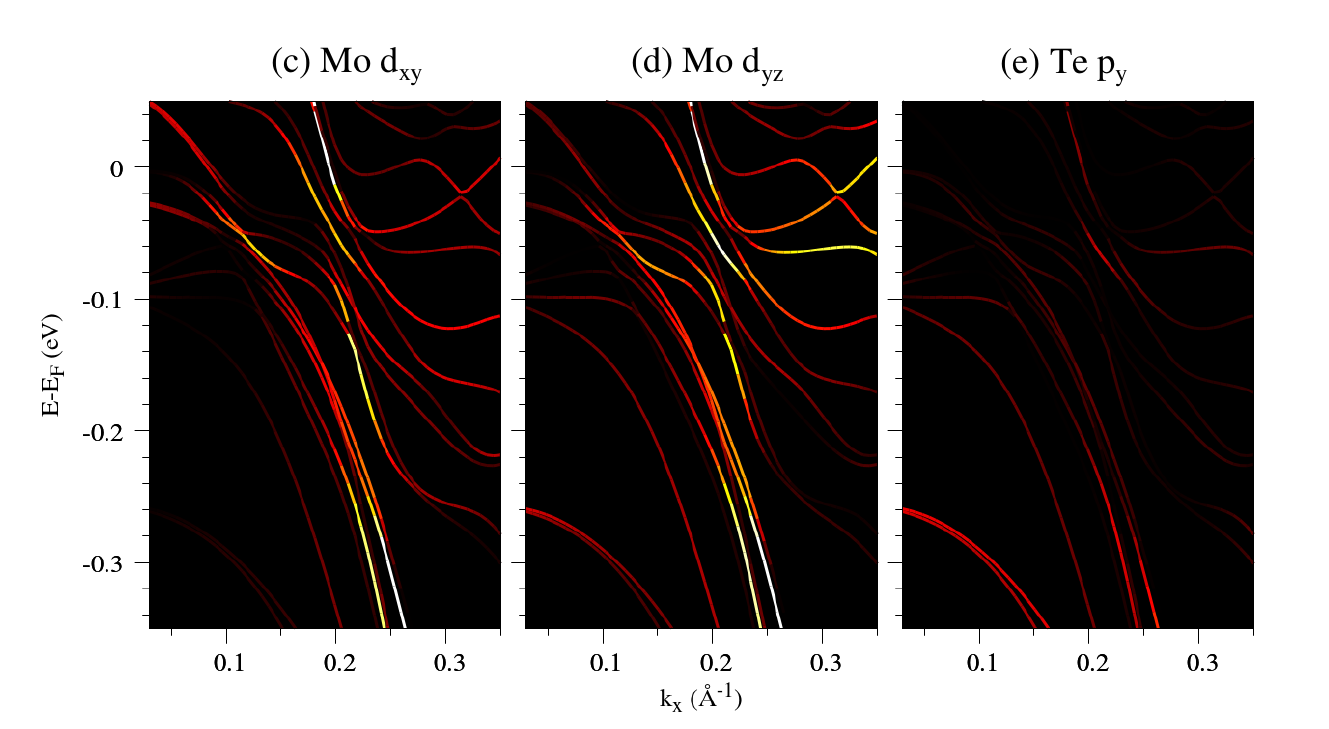}
\caption{\label{fig:expe}
Comparison between experimental (a) and theoretical (b) ARPES intensity maps
for $\hbar \omega = 6.27$~eV and s-polarized light.
(c-e) Corresponding orbital projected bands. This is the same data
as in Fig.~\ref{fig:ypol} plotted on an enlarged $E$-scale for easy comparison with (a,b).
}
\end{figure}

Here we have measured the ARPES of MoTe$_2$ near the Fermi energy
along the $\bar{\Gamma}$-$\bar{X}$ line ($k_x<0.4$~\AA$^{-1}$)
using a photon energy of 6.27~eV and $s$-polarized light
(polarization vector along the $y$-axis).
See Fig.~1~c for the experimental geometry.
The experimental data is shown in
Fig.~\ref{fig:expe}, along with the corresponding one-step ARPES calculation.
Here, the raw ARPES intensity is shown. The second derivative of the intensity
with respect to the energy is often used to enhance the dispersion of broad
spectra features. The corresponding experimental data is shown in Fig.~S6
of the Supplemental Material~\cite{SI}.
Both the measured and the calculated ARPES map are
dominated by a group of strongly dispersing bands, marked 2-4
in Fig.~\ref{fig:expe}~a. 
As seen from the comparison with the projected bands (Fig.~\ref{fig:expe}, c-e)
these features are mainly of Mo~$d_{xy}$ and Mo~$d_{yz}$ character.
Feature~4 may be assigned to the hole pocket while features 1-3 belong
to lower energy bands with similar dispersion. 
Features 5 and 6 are part of the electron pocket,
where the intense band~5, is dominated by Mo~$d_{xy}$ and $d_{yz}$ orbitals, while
the weak feature~6 is essentially of Mo~$d_{yz}$ character.
Experiment and theory disagree somewhat about the $k_x$-positions of the
various features. The calculated feature~2 is shifted by 0.1~\AA$^{-1}$
to higher $k_x$ with respect to experiment,
which could be due to limitations of the surface barrier model used in the
KKR calculations.
Furthermore, the splitting between hole and electron pocket is smaller in the
calculation than in experiment (features 4 and 5). 
Apart from these details, the calculated ARPES intensity map agrees
very well with the data.
 
The ARPES calculations in Fig.~\ref{fig:ypol}~a and Fig.~\ref{fig:expe}~b were obtained with
the same light polarization but different photon energies,
60 and 6.27~eV, respectively.
When comparing the same binding energy range, we find that the two ARPES
intensity maps are very similar, except that with 6.27~eV photons, the average
intensity is over one order of magnitude larger than with 60~eV photons.

\subsection{Photon energy and $k_z$ dependence}
\begin{figure}[htb]
\centering
 \includegraphics[width=17cm]{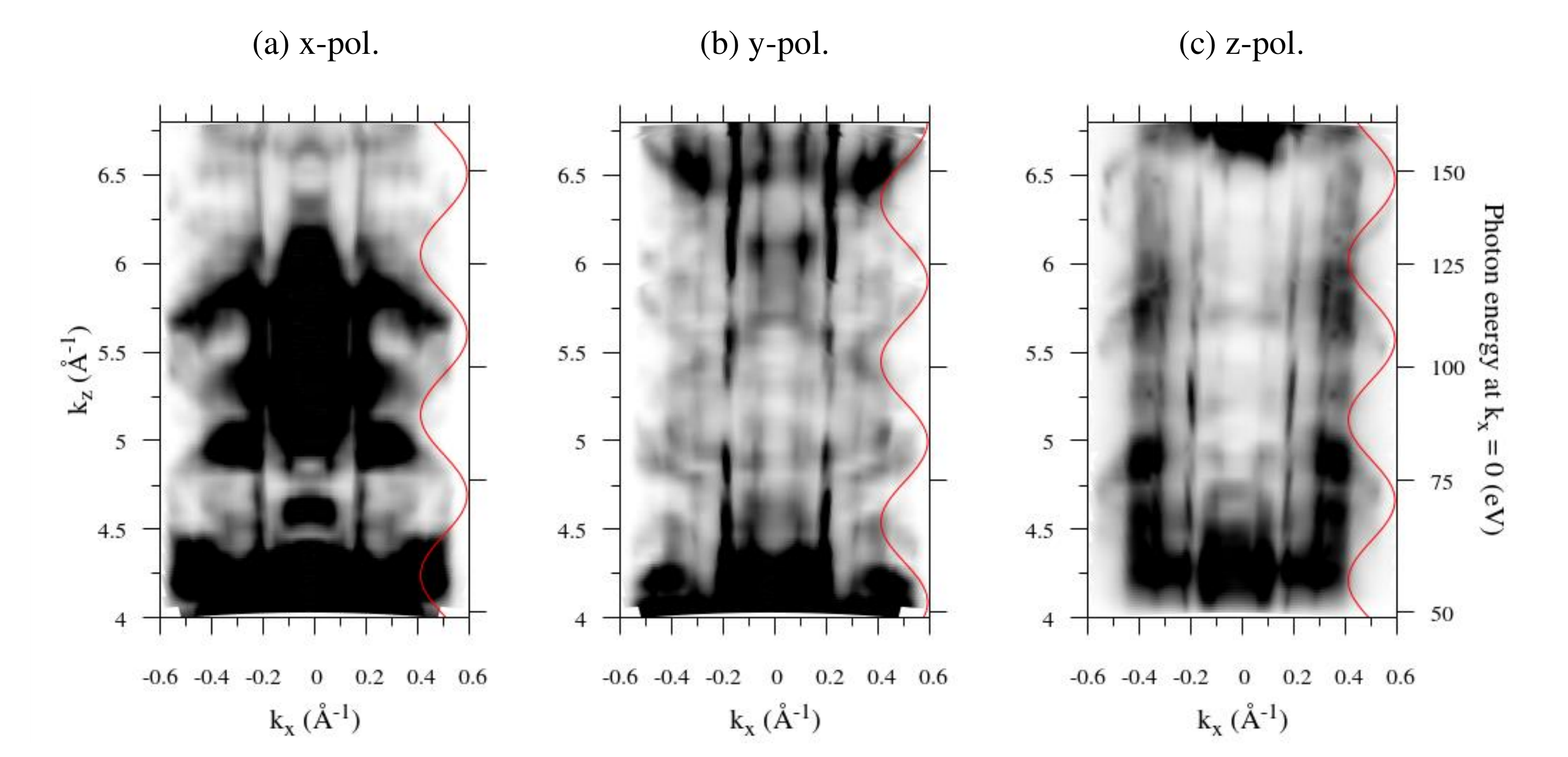}
\caption{\label{fig:kz}
k$_x$-k$_z$ dispersion of the ARPES intensity calculated from one-step model for x-polarization (a), y-polarization (b) and z-polarization (c). The red solid line is a cosine function with period $4\pi/c \approx 0.905$~\AA$^{-1}$.}
\end{figure}

We have computed the ARPES from the Fermi level along the $\bar{\Gamma}$-$\bar{X}$ line,
as a function of photon energy in the energy range
50-160 eV. In the spirit of the three-step model, the photon energy
has been converted to the final state photoelectron momentum $k_z$
in the bulk by using $k_z=\sqrt{2m(E_{\rm kin} +V_0)/\hbar^2-k_{||}^2}$.
Here $E_{\rm kin}$ is the kinetic energy of photoelectron and $V_0=16$~eV is the inner potential.
The value of $V_0$ has been estimated from the work function $\phi=4.1$~eV and the
KKR interstitial potential and is comparable with Ref.~\cite{Tamai2016}.
The ARPES map intensity (Fig.\ref{fig:kz}) shows a strong dependence on $k_z$.
The most intense bands have no $k_z$ dispersion, indicating surface
band character.
However, especially the region $0.4$~\AA$^{-1}$ $< |k_x| < 0.6 $~\AA$^{-1}$
displays a clear period of $4\pi/c \approx 0.905$~\AA$^{-1}$.
The $4\pi/c$ periodicity is in agreement with the data of Ref.~\cite{Tamai2016}
and was also observed in WTe$_2$~\cite{Disante2017}.
Our one-step calculations thus prove theoretically that some of the $k_z$ dispersion of the
MoTe$_2$ bulk band structure survives in the photoemission process and can be
observed in ARPES. Interestingly the period of the $k_z$ oscillation is
$4\pi/c$ rather than $2\pi/c$, the value expected from the band structure.
The reason is that MoTe$_2$ has a non-symmorphic space group with glide planes
along~$c$. As shown by Pescia et al~\cite{Pescia1985}, for such space groups,
final states change parity with respect to the glide
plane when going from $k_z$ to $k_z+2\pi/c$.
As a consequence photoemission transition is possible only for either of
the two parity related initial state bands.
This effect has been observed in other layered materials including graphite~\cite{Pescia1985,Matsui2018} and MoTe$_2$ (2H)~\cite{Boker2001}.

\section{\label{sec: conclusions} Conclusions}
In summary, we have presented a combined experimental and theoretical
study of the band structure of the Weyl semi-metal candidate material
MoTe$_2$(T$_d$).  The orbital character of the near Fermi-energy bands has
been analyzed using density functional theory. We find that along the
$\bar{\Gamma}-\bar{X}$ line, the hole pocket is dominated by the Te~$p_x$ orbital,
while the electron pocket is mainly made of Mo~$d_{yz}$ and Te~$p_z$ derived
bands.  We have performed one-step-model ARPES calculations and
obtained good agreement with the experimental data. 
The ARPES intensity depends strongly on the light
polarization, which can be understood from the orbital character of
the bands together with the dipole selection rules. The results show
how relevant parts of the near Fermi-energy band structure, especially
electron-pocket, hole-pocket and surface state, can be highlighted using the most
suitable light polarization. 
The calculated ARPES maps have a complex photon energy dependence and
display an approximate $4\pi/c$ periodicity in $k_z$, in agreement with experiment.
More generally, we conclude that one-step-model ARPES calculations are
indispensable for a quantitative interpretation of ARPES data with
complex band structures typical for topological 2D materials.

\section{\label{sec: Ac} Acknowledgments}
We are very grateful to Prof. Keiji Ueno of Saitama University for providing us the MoTe$_2$ sample.
R.O. thanks financial support by the NIM Summer Research Program of
the University of Munich (LMU)
and by the Frontier Science Program of Chiba University.
J.S. and J.M. would like to thank CEDAMNF project financed by the Ministry
of Education, Youth and Sports of Czech Republic,
Project No. CZ.02.1.01/0.0/0.0/15$\_$003/0000358.
J.B. and H.E. acknowledge financial support by the DFG via the projects
Eb 158/32 and Eb 158/36.
\bibliographystyle{apsrev4-2}
\bibliography{library}


\end{document}


\title{Supplemental Material \\
for \\
Surface band characters of Weyl semimetal candidate material
MoTe$_2$ revealed by one-step ARPES theory}

\author{Ryota Ono}
\affiliation{Graduate School of Science and Engineering, Chiba University, Inage-ku, Chiba-shi 265-8522, Japan}
\author{Alberto Marmodoro}
\affiliation{FZU - Institute of Physics of the Czech Academy of Sciences, Cukrovarnicka 10, CZ-162 53 Prague, Czech Republic}
\author{Jakub Schusser}
\affiliation{New Technologies - Research Center, University of West Bohemia, Univerzitni 8, 306 14 Plze\v{n}, Czech Republic}
\affiliation{Experimentelle Physik VII, Universit\"at W\"urzburg, Am Hubland, D-97074 W\"urzburg, Germany}
\author{Yositaka Nakata}
\affiliation{Graduate School of Science and Engineering, Chiba University, Inage-ku, Chiba-shi 265-8522, Japan}
\author{Eike F. Schwier}
\affiliation{Hiroshima Synchrotron Radiation Center, Hiroshima University, Kagamiyama 2-313, Higashi-Hiroshima 739-0046, Japan}
\affiliation{Experimentelle Physik VII, Universit\"at W\"urzburg, Am Hubland, D-97074 W\"urzburg, Germany}
\author{J\"urgen Braun}
\affiliation{Department of Chemistry, Ludwig Maximililans University M\"unchen, Butenandtstra\ss e 11, 81377 M\"unchen, Germany}
\author{Hubert Ebert}
\affiliation{Department of Chemistry, Ludwig Maximililans University M\"unchen, Butenandtstra\ss e 11, 81377 M\"unchen, Germany}
\author{J\'{a}n Min\'{a}r}
\affiliation{New Technologies - Research Center, University of West Bohemia, Univerzitni 8, 306 14 Plze\v{n}, Czech Republic}
\author{Kazuyuki Sakamoto}
\affiliation{Graduate School of Engineering and Molecular Chirality Research Center, Chiba University, Chiba 263-8522, Japan}
\affiliation{Department of Applied Physics, Osaka University, Osaka 565-0871, Japan}
\author{Peter Kr\"uger}
\email{pkruger@chiba-u.jp}
\affiliation{Graduate School of Engineering and Molecular Chirality Research Center, Chiba University, Chiba 263-8522 Japan}

\date{\today}

\date{\today}

\maketitle
\newpage
\section{\label{sec:pdos} Density of states around the Fermi level}
Here we show that the bands close to the Fermi level are dominated by
the Mo-$d$ orbital and the Te-$p$ orbital as we indicated in the main text.

The density of states around the Fermi level, projected on the surface (Te) and sub-surface atoms (Mo) is shown in Fig.~\ref{fig: pdos}.
As one can see, the Mo-$d$ orbital and the Te-$p$ orbital are dominant 
close to the Fermi level, while there is only a small contribution of the other orbitals.

\begin{figure}[htb]
\centering
 \includegraphics[width=9cm]{Plots/pdos.pdf}
        \caption{\label{fig: pdos} Comparison of the density of states around the Fermi level 
        projected onto the orbitals of the first (Te) and second (Mo) atomic layer.}
\end{figure}

\section{\label{sec:other} Projected band structures}
In this section, we show all projected band structures (up to $l=2$) in order to justify our analysis in Sec.~2 in the main text.

\begin{figure}[htb]
 \centering
 \begin{tabular}{ccc}
 \includegraphics[width=4.5cm]{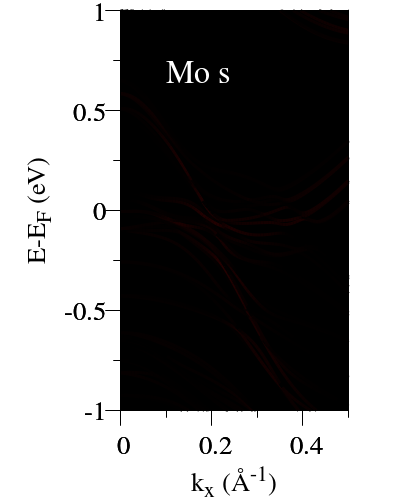} &
 \includegraphics[width=4.5cm]{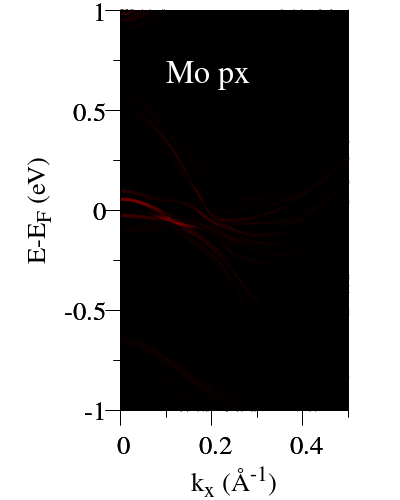} &
 \includegraphics[width=4.5cm]{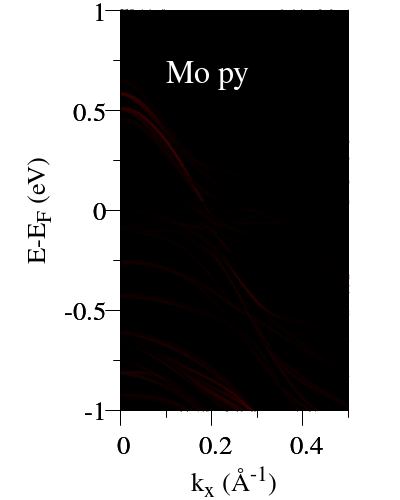} \\ 
 \includegraphics[width=4.5cm]{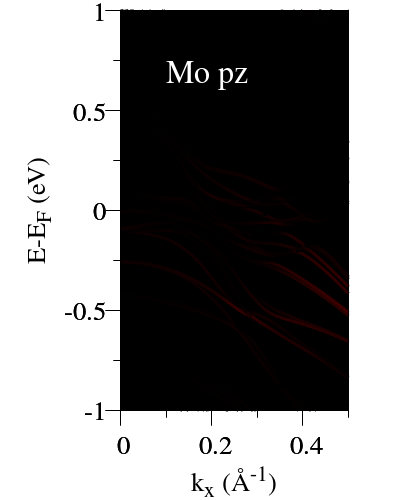} &
 \includegraphics[width=4.5cm]{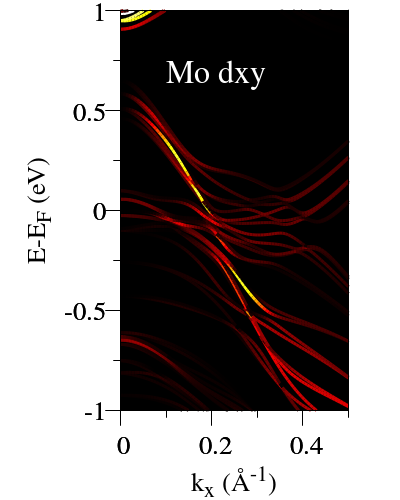} &
 \includegraphics[width=4.5cm]{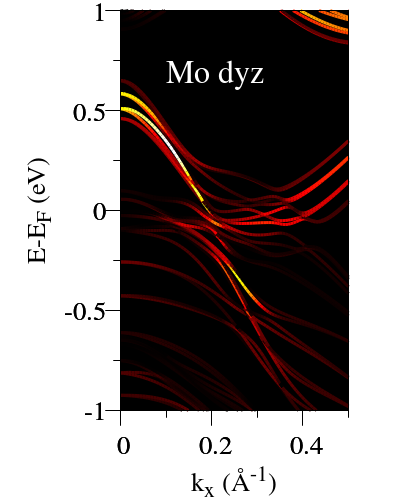} \\ 
 \includegraphics[width=4.5cm]{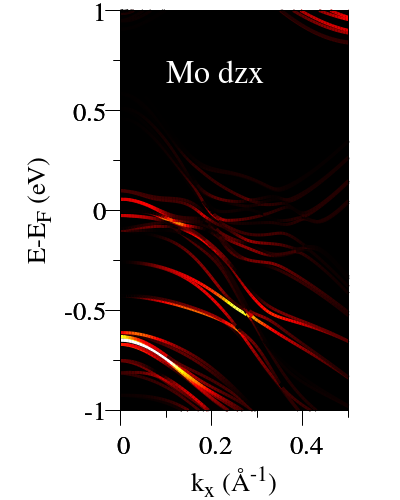} &
 \includegraphics[width=4.5cm]{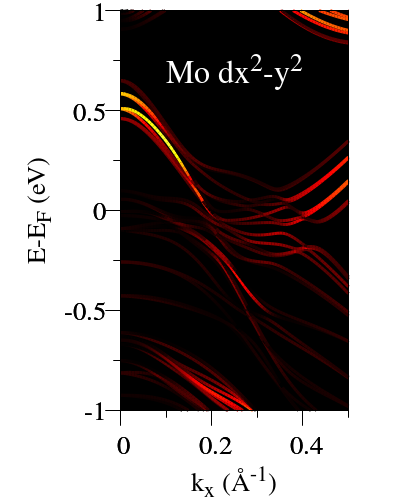} &
 \includegraphics[width=4.5cm]{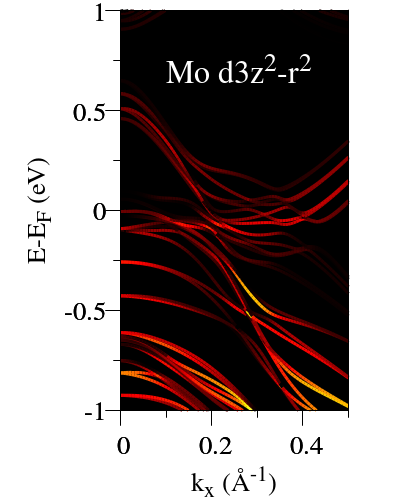} \\ 
 \end{tabular}
        \caption{\label{fig: Mo_pbands} DFT bands projected on Mo orbitals with $l=0, 1, 2$.}
\end{figure}

\begin{figure}[htb]
 \centering
 \begin{tabular}{ccc}
 \includegraphics[width=4.5cm]{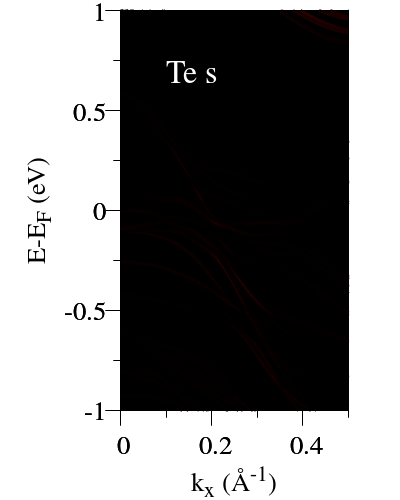} &
 \includegraphics[width=4.5cm]{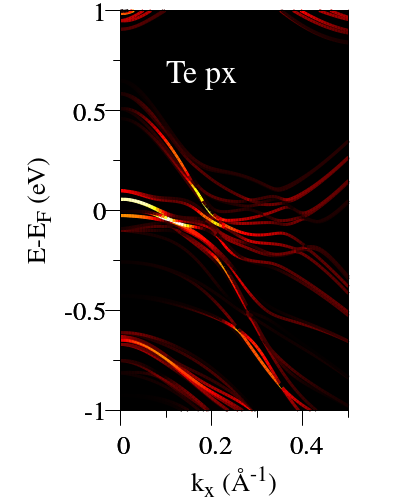} &
 \includegraphics[width=4.5cm]{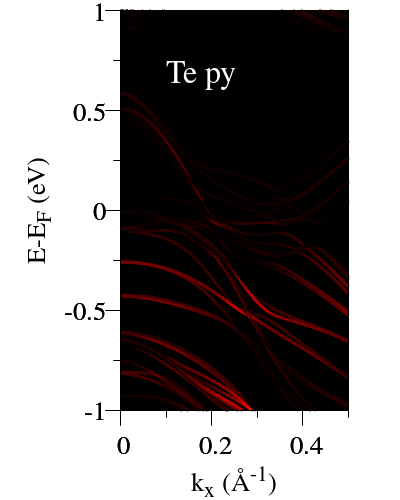} \\ 
 \includegraphics[width=4.5cm]{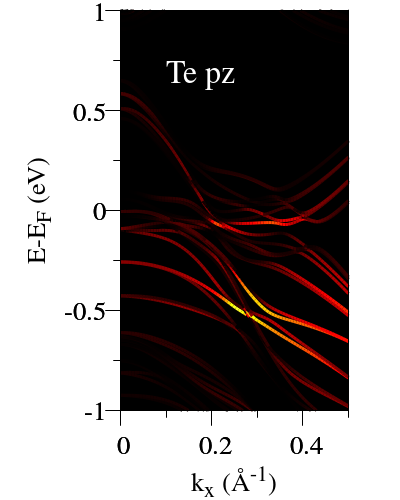} &
 \includegraphics[width=4.5cm]{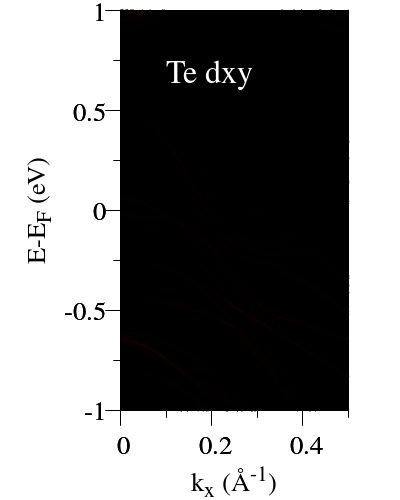} &
 \includegraphics[width=4.5cm]{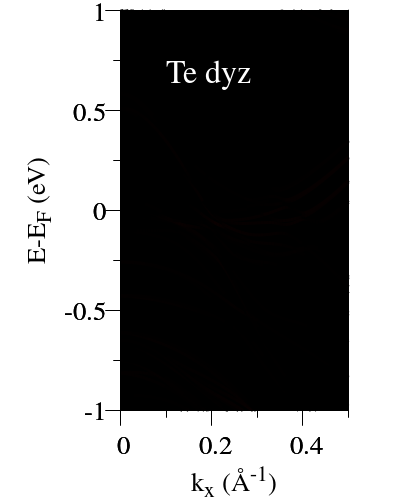} \\ 
 \includegraphics[width=4.5cm]{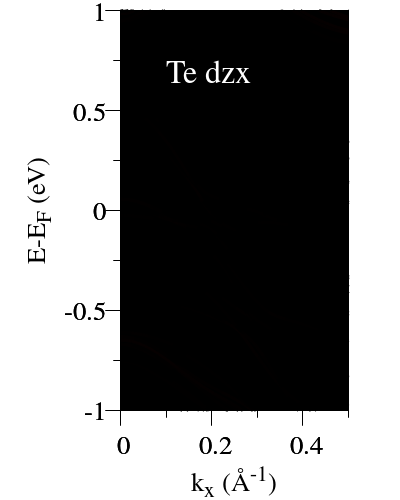} &
 \includegraphics[width=4.5cm]{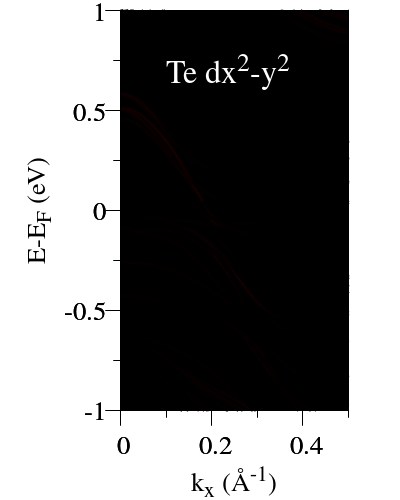} &
 \includegraphics[width=4.5cm]{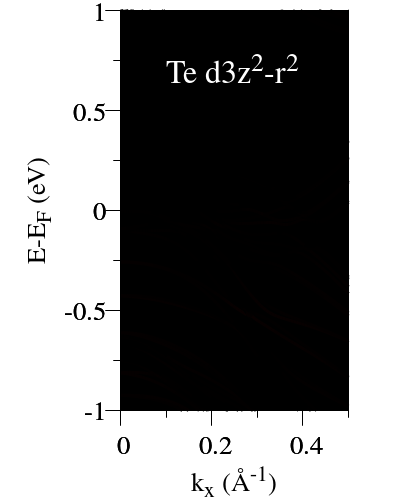} \\ 
 \end{tabular}
        \caption{\label{fig: Te_pbands} DFT bands projected on Te orbitals with $l=0, 1, 2$.}
\end{figure}

Figs.~\ref{fig: Mo_pbands} and ~\ref{fig: Te_pbands} show the band structures projected onto all the orbitals 
of the first layer Mo atoms and Te atoms, respectively. As in the main text, the Mo intensity is multiplied by
a factor~0.5.
It can be seen that the orbital weights of Mo~s,p as well as Te~s,d eigenstates are negligible.
Moreover, the Te-p and Mo-$d$ projections that are {\em not shown}\/ in the main text, do not fit well the ARPES intensity maps, 
which justifies our analysis of Sec.~2.B and 2.C.

\section{\label{sec:Ecut} ARPES intensity at Fermi energy}
\begin{figure}[htb]
 \centering
 \includegraphics[width=10cm]{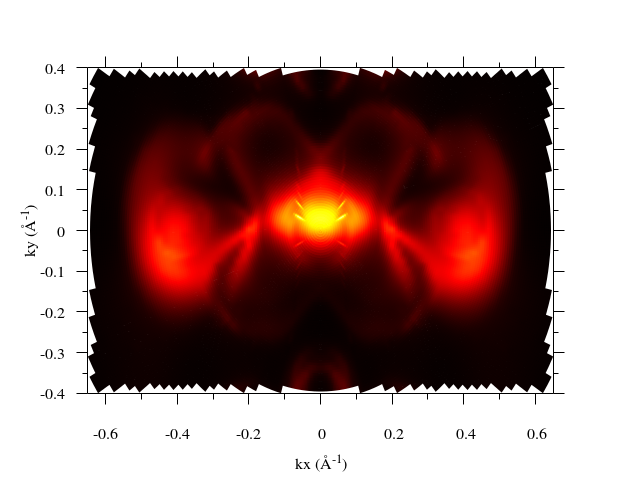}
 \includegraphics[width=10cm]{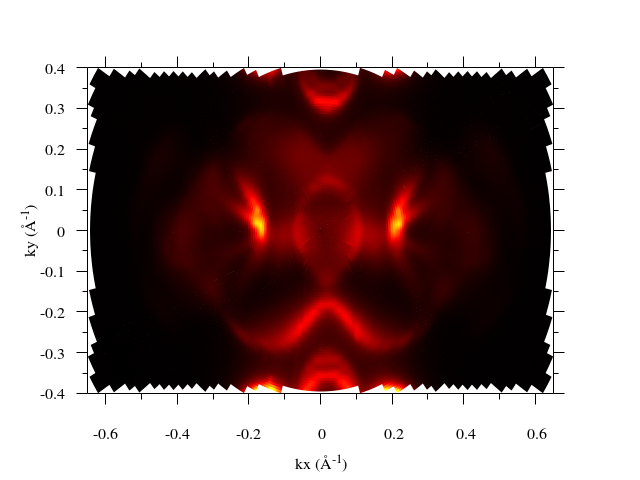}
 \includegraphics[width=10cm]{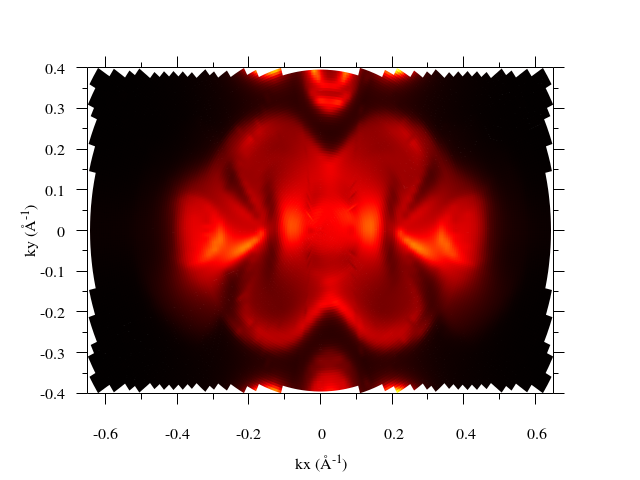}
        \caption{\label{fig:contour} ARPES intensity at the Fermi energy for x-polarization (top), y-polarization (middle) and
                 z-polarization (bottom) with photon energy $\hbar \omega = $60~eV. The color scale is same as in Figs~3-5. }
\end{figure}

Fig.~\ref{fig:contour} shows theoretical ARPES intensity at the Fermi energy for x, y and z-polarization with photon energy $\hbar \omega = $60~eV.
Different features can be highlighted by those different polarizations. 
The map with z-polarization is in good agreement with the data of Ref.~\cite{Tamai2016} where $p$-polarized light was used.

\section{\label{sec:sp_ARPES} Theoretical spin-polarized ARPES}
\begin{figure}[htb]
 \centering
 \includegraphics[width=17cm]{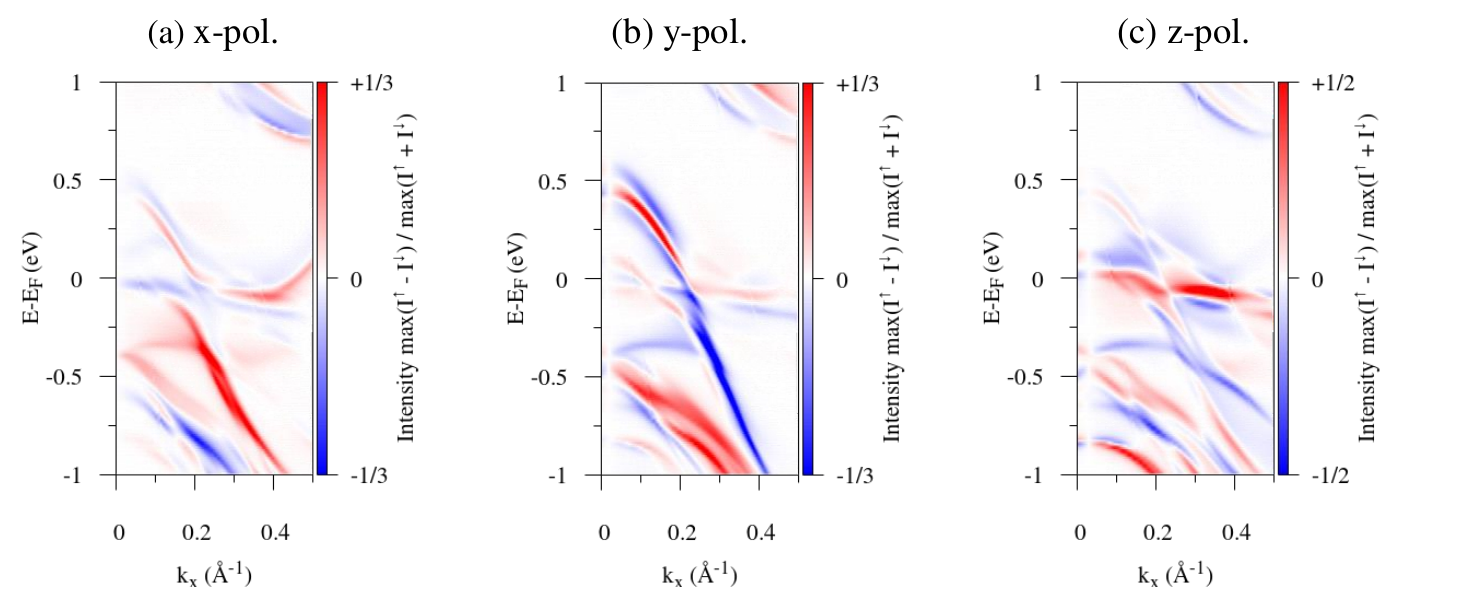}
        \caption{\label{fig:sp_ARPES} Theoretical spin-polarized ARPES intensity map with x (a), y (b) and z-polarization (c) with photon energy $\hbar \omega = $60~eV.
                                      The spin-quantization axis is taken as the z-axis. 
                                      The plots are corresponding to the ARPES of Figs.~3, 4 and 5 of the main text. }
\end{figure}

Fig.~\ref{fig:sp_ARPES} shows theoretical spin-polarized ARPES with x, y and z-light polarization with photon energy $\hbar \omega = $60~eV. 

Because of space-inversion symmetry breaking and a large spin-orbit coupling,
the bands of MoTe$_2$ have a complex spin-texture~\cite{Weber2018}.
In our DFT calculations we found
that the MoTe$_2$ bands are spin-orbit split up to 0.05~eV at certain k-points
and have appreciable spin-polarization in agreement with the literature~\cite{Weber2018,Crepaldi2017}
This give rise to spin-polarization of the photoelectrons which has
has been confirmed experimentally for the Fermi-surface~\cite{Weber2018}.
However, the interplay between light polarization and photoelectron
spin-polarization has not been studied yet.
It can be seen that many ARPES bands are strongly spin-polarized.
Often spin-up band is accompanied by a spin-down band next to it,
with a band splitting of the order of 0.1~eV,
reflecting the spin-orbit splitting of the initial bands.
Interestingly, the intense surface band at $E \approx 0$~eV, $k_z= 0.3-0.4$~\AA$^{-1}$
has also a large spin-polarization which can reach 50\%
with z-pol light (Fig.~\ref{fig:sp_ARPES}~c).
 
\section{\label{sec:2nd} Second derivative of the ARPES intensity}
\begin{figure}[htb]
 \centering
 \includegraphics[width=6.5cm]{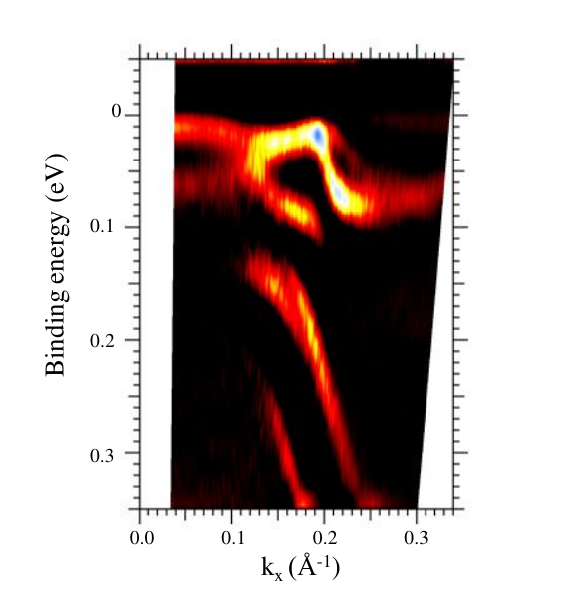}
        \caption{\label{fig:2nd} Second-derivative of the ARPES intensity map.}
\end{figure}

To visualize the characters of the ARPES intensity, we show the second derivative
of the experimental ARPES intensity in Fig.~\ref{fig:2nd}.  The band dispersion is highlighted more clearly in this picture. 
However, for direct comparison with the theory, one needs to look at the raw ARPES intensity data
shown in the Sec.~3 of the main text.
 
\bibliographystyle{apsrev4-2}
\bibliography{library}